\RequirePackage{fix-cm}
\documentclass[twocolumn]{svjour3}          
\smartqed  
\usepackage{graphicx}
\usepackage{multicol}
\usepackage[labelfont=bf]{caption}
\usepackage[colorlinks = true,
linkcolor = blue,
urlcolor  = blue,
citecolor = blue,
anchorcolor = blue]{hyperref}
\usepackage{url}
\usepackage{amsmath}
\usepackage{ulem}

\begin{document}

\title{Effect of wall friction on 2D hopper flow
}



\author{Neil Shah               \and
        Brenda Carballo-Ramirez \and
        Sumit Kumar Birwa       \and
        Nalini Easwar           \and
        Shubha Tewari           \and
}

 \institute{Neil Shah \at
Department of Physics and Astronomy, Tufts University, 574 Boston Ave., Medford MA
    \\\email{neil.shah@tufts.edu}         
    \and
    Brenda Carballo-Ramirez 
    \and Nalini Easwar
    \at
    Department of Physics, Smith College, Northampton MA 01063
    \\\email{neaswar@smith.edu}
    \and
    Sumit Kumar Birwa
    \at Department of Applied Mathematics and Theoretical Physics, University of Cambridge, Cambridge CB3 0WA, United Kingdom
    \at International Centre for Theoretical Sciences, Tata Institute of Fundamental Research, Shivakote, Bengaluru 560089, India
    \and Shubha Tewari
    \at Department of Physics, University of Massachusetts, Amherst, MA 01003
    \\\email{tewari@umass.edu}
 }

\date{Received: date / Accepted: date}

\maketitle

\begin{abstract}

We report here on experiments and simulations examining the effect of changing wall friction on the gravity-driven flow of spherical particles in a vertical hopper. In 2D experiments and simulations, we observe that the exponent of the expected power-law scaling of mass flow rate with opening size (known as Beverloo's law) {\it decreases} as the coefficient of friction between particles and wall increases, whereas Beverloo scaling works as expected in 3D. In our 2D experiments, we find that wall friction plays the biggest role in a region near the outlet comparable in height to the largest opening size. However, wall friction is not the only factor determining a constant rate of flow, as we observe a near-constant mass outflow rate in the 2D simulations even when wall friction is set to zero. We show in our simulations that an increase in wall friction leaves packing fractions relatively unchanged, while average particle velocities become independent of opening size as the coefficient of friction increases. We track the spatial pattern of time-averaged particle velocities and accelerations inside the hopper. We observe that the hemisphere-like region above the opening where particles begin to accelerate is largely independent of opening size at finite wall friction. However, the magnitude of particle accelerations decreases significantly as wall friction increases, which in turn results in mean sphere velocities that no longer scale with opening size, consistent with our observations of mass flow rate scaling. The case of zero wall friction is anomalous, in that most of the acceleration takes place near the outlet. \par

\end{abstract}

\newpage

\section{Introduction}
\label{intro}

The flow of granular materials out of a container under gravity has long been a topic of both scientific and practical interest. In this paper, we focus on an age-old observation: the rate of flow of granular materials through a vertical hopper is constant, independent of column height, until the column is almost fully drained \cite{Nedderman1982Flow}. Repeated measurements have shown that this constant mass flow rate scales as the opening size to a characteristic power that depends on the dimensionality of the system, a phenomenon labeled Beverloo's law \cite{Beverloo1961Flow}. We perform experiments and simulations in a quasi-2D hopper that examine the effect of wall friction on gravity-driven mass flow rates. We find that changing wall friction alters the effective scaling exponent of the power-law dependence of mass flow rate on opening size. \par

At first glance, the idea that wall friction alters how the mass efflux scales with opening size does not make sense, since the scaling power law is usually cast in terms of dimensional arguments: $W \propto a^{5/2} (a^{3/2})$ in three (two) dimensions, where $a$ is the diameter of the opening. A power of 2 in 3D (1 in 2D) comes from the area (length) of the opening, and the additional $1/2$ power from the idea that there is a ‘free-fall’ region of height equivalent to the opening diameter $a$, allowing particles to gain speed proportional to $a^{1/2}$ as they exit, an idea first proposed by Hagen in 1852 \cite{Tighe2007Pressure}. Given that the flow rate vanishes for a finite opening size, Beverloo \cite{Beverloo1961Flow} introduced a correction to the scaling law that accounts for the finite size of the particles as they crowd the opening, i.e. $$W = W_0 (a - kd)^{3\over 2}$$ in 2D where $W_0$ and $k$ are experimentally determined fitting parameters and $d$ is the particle diameter \cite{Beverloo1961Flow}. The free-fall region idea was later refined \cite{Hilton2011Granular} to an assumption of a hemispherical arch that marks the boundary between the region of dense flow and a more dilute region in which particles lose contact and begin to free-fall. This assumption allowed a derivation of the Beverloo expression for the flow rate, with a proportionality constant that is purely geometry-dependent. \par

There is no clear experimental evidence for the formation of a physical free-fall arch. Experiments that track particles and contact forces using photoelastic disks do observe that these arch-like regions occur intermittently, but only for small openings, and find no direct connection between arch formation and outflow rate \cite{Vivanco2012Dynamical}. Rather, it appears from experiments and simulations that there is a change in kinetic pressure over a region that scales with the opening size \cite{Rubio2015Disentangling}, and the effective acceleration of the exiting particles increases over this same region in a fashion that also scales with the opening size. The proximity of the walls also doesn't appear important - experiments on granular hoppers have shown that that the mass flow rate is purely a function of the opening size, independent of the width of the hopper, as long as the width is at least 2.5 times the size of the outlet \cite{NeddermanBook1992}. \par

We report here that our 2D experiments and simulations indicate, consistent with these previous observations, that there is an acceleration region near the opening. The width of this region (in the horizontal direction) grows with the size of the opening, but there is little accompanying change in the height above the opening at which particles first begin to accelerate. While both our experiments and simulations indicate that friction at the side walls plays little to no role, we find that increasing friction on the front and back walls of the quasi-2D hopper strongly impedes the extent to which particles are able to accelerate. As a result, for high wall friction, sphere velocities at the outlet no longer scale with opening size, in apparent violation of Beverloo's law. We emphasize, though, that this appears to come from the loss of kinetic energy due to particle collisions with the front and back walls. Significantly, our 2D experiments find no difference in the flow rates for hoppers fully lined with a particular frictional backing and hoppers in which only a region near the opening has that frictional backing. This means the flow rate is determined almost entirely by the friction in a region near the outlet. In 3D experiments, however, we observe no significant effect of wall friction on how the flow rate scales with opening size \cite{Smith}, and Beverloo's law holds. \par

While the presence of frictional forces is crucial for achieving a constant rate of flow, the precise form of the frictional interactions needed, as well as the role of the walls, is less clear. Also not clear is whether these frictional forces are required to create a condition of constant pressure at the outlet while grains are flowing out of the hopper. The well-known observation that grains flow at a constant rate out of a hopper is in stark contrast with a column of liquid, in which energy conservation dictates that the outflow rate at the base decreases with the column height because of the lower potential energy available to the system. The usual explanation for constant granular flow relies on the assumption of an unchanging net potential energy at the base, arising from an effective column weight that becomes independent of column height. It has long been observed that the pressure at the base of a {\it static column} of increasing height saturates over a characteristic length comparable to the column width, a phenomenon called the Janssen effect \cite{Janssen1895}. The idea is that when this happens, the excess weight of added grains is distributed laterally rather than downwards, balanced by the upward frictional force at the walls. The Janssen effect is also found when grains are in motion relative to the walls \cite{Bertho2003Dynamical}, though the apparent weight saturates at values lower than in the static case. However, measurements of pressure near the outlet in a system of disks moving at constant speed on a horizontal conveyor belt \cite{Aguirre2010} found that pressure at the outlet and the outflow speed were decoupled -- one could be made to vary while the other was held constant. We report here that, consistent with this last observation, and in contrast to what we had expected for vertical gravity-driven flow, our simulations indicate a near-constant flow rate even when the particle-wall frictional coefficient is set to zero. We do retain inter-particle friction in these cases, and our results appear to indicate that the mechanism for slowing down particles is dynamic in nature. However, what happens to the dynamic pressure in the vicinity of the outlet as wall friction changes remains to be investigated. \par

In what follows, we first describe our experiment and simulation method, and then present results for how flow rates change with wall friction and opening size for both experiment and simulation in 2D. We then use our simulations to explore the dynamics further by reporting on the spatial pattern of time-averaged particle velocities and accelerations inside the hopper as wall friction and the opening size at the base are changed. \par

\section{Methods}
\label{methods}

\subsection{Experiment}

\begin{figure*}
\begin{center}
  \includegraphics[width=0.75\textwidth]{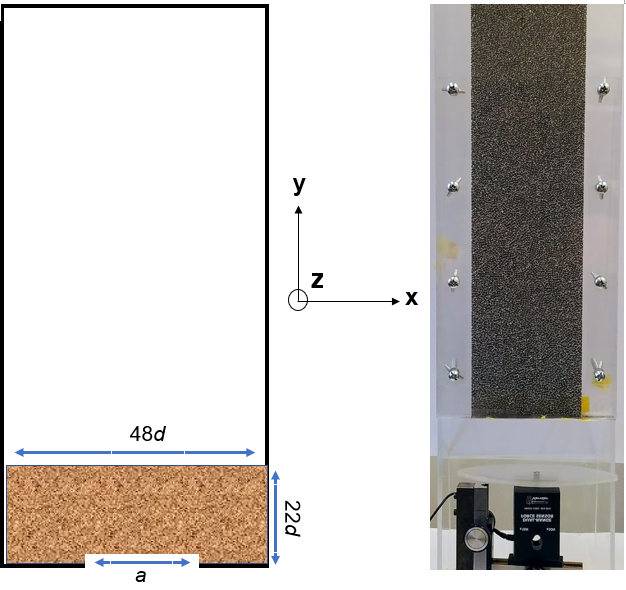}
\caption{(Left) Schematic of experimental hopper with the OS/BP (Opening Sandpaper, Bulk Paper) lining: the bulk is lined with paper, with the rough region of height 22d near the opening lined with sandpaper. (Right) Photograph of experimental hopper of dimensions 400 x 48 x 1.2 in units of particle diameter, fully lined with sandpaper. The hopper is filled with steel spheres of diameter $\approx 2.5$ mm, which flow out of an opening at the base onto a box sitting on a force transducer that records the weight as a function of time. The slope of this force trace is a measure of the mass flow rate. We control the friction on the walls by changing the material backing on the front and back walls. \\
}
\label{Schematic}
\end{center}
\end{figure*}

\begin{figure*}
\begin{center}
  \includegraphics[width=1.0\textwidth]{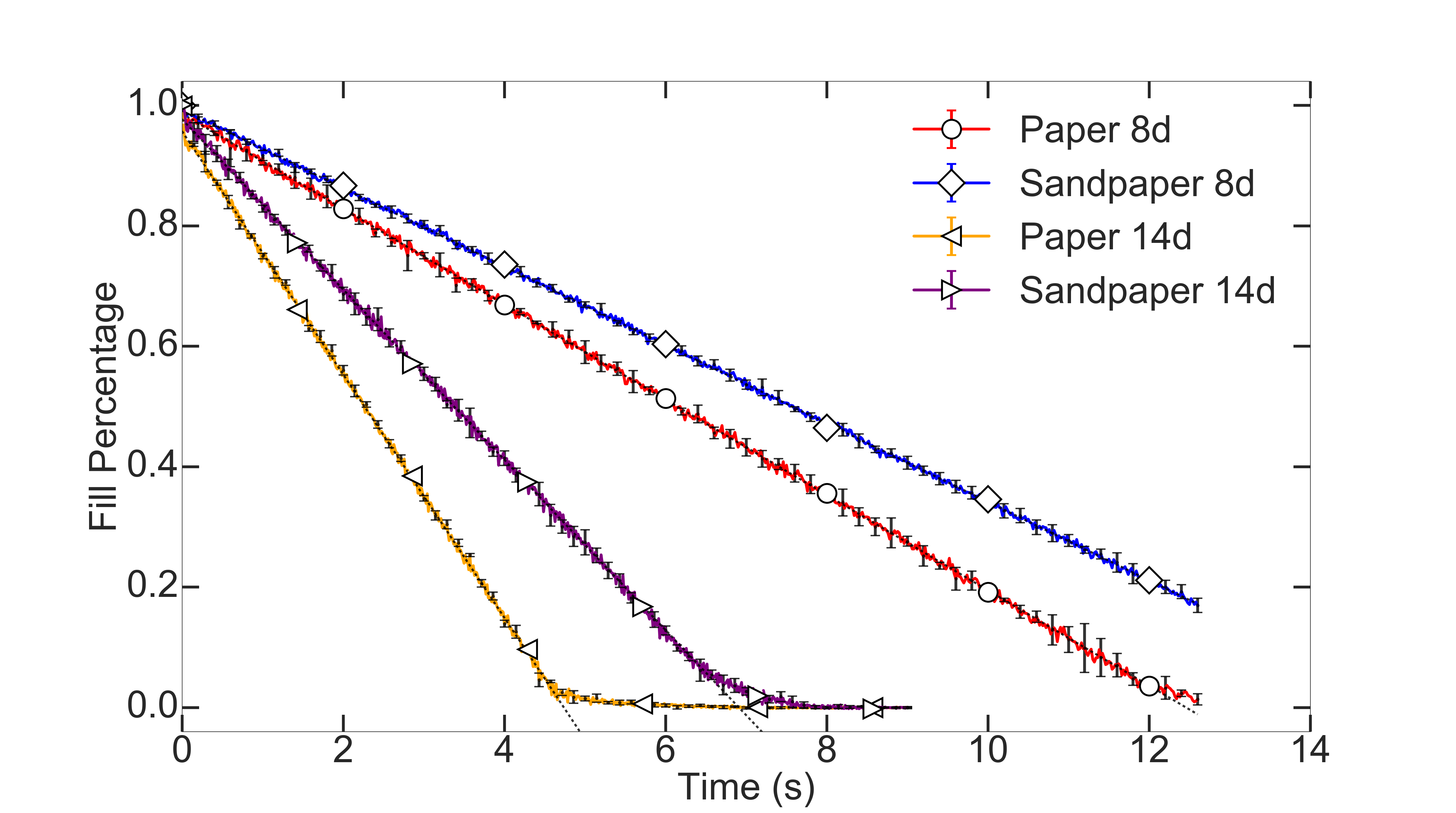}
\caption{Experimental fill percentage as a function of time for a draining hopper at two representative opening sizes (8$d$ and 14$d$) and for two distinct wall backings: paper and sandpaper. Each curve is averaged over five experimental runs, with error bars indicating the mean deviation over these runs. The fill percentage is the ratio of the height of the granular pile at a given time to the initial fill height and is extracted from the force trace data. Error bars are shown every 10 data points for better visibility. The dashed lines represent straight line fits to the data.}
\label{ExperimentFlows}
\end{center}
\end{figure*}

The experimental setup is a quasi-2D rectangular prism 48 particle diameters ($d$) wide, 400$d$ tall and 1.2$d$ in depth. (All length measurements are expressed in particle diameters $d$). The front and back walls are made of plexiglass and are lined with different frictional backings. The three materials used are Teflon, poster-board (paper), and 60-grit sandpaper. The right-hand side of Fig.~\ref{Schematic} shows the experimental setup with the sandpaper backing. To study where wall friction plays the biggest role in the hopper, we created hybrid linings, see left hand-side of Fig.~\ref{Schematic}, in which one material was used for the bulk of the lining (white region), and the other material (shaded region) was used to line a region of height 22$d$ above the opening. There are two realizations of these hybrid linings: sandpaper at the opening and paper in the bulk (referred to as OS/BP) and the reverse, OP/BS. The grains are steel spheres, 2.5 mm in diameter, thus large enough to neglect the effects of air resistance. The 1.2$d$ depth of the hopper ensures that the particles form a single layer and undergo significant collisions with the front and back walls. The grains are filled to the same height for all the wall frictions and openings, then allowed to flow through a rectangular opening at the base into a container. The opening size $a$ is varied from 6$d$ $\rightarrow$ 22$d$ in increments of 2$d$. A force sensor placed beneath the outlet container allows a measurement of the accumulated weight as a function of time.  This measured weight vs time data, converted to a fill percentage of the draining hopper for better comparison with simulation, is plotted as a function of time for two different opening sizes and two different hopper linings in Fig.~\ref{ExperimentFlows}. The data for each opening and wall backing are averaged over five experimental runs, with the corresponding error bars shown on the figure. The straight line fits (dashed lines) show that the flow rates remain  linear till the hopper is almost fully drained, and this flow rate is smaller for a given opening when the wall backing is more frictional, as might be expected. Doing linear fits to the weight vs time data for all five hopper linings and all nine openings, we are able to look at how flow rate scales with opening size as wall friction changes. The results for all five hopper linings, plotted using open symbols in Fig.~\ref{FlowvsOpening}, indicate that the power law exponent decreases as friction increases. We thus use simulations to try and understand this phenomenon better. \par

\subsection{Simulations}

We set up our simulations using the granular package of LAMMPS \cite{LAMMPS}. The simulation geometry is chosen to closely match the specifications of the experimental setup in Fig.~\ref{Schematic}: a quasi-2D rectangular box with a rectangular opening at the base, filled with monodisperse spheres. Length and mass units are expressed in units of sphere diameter and mass, and the acceleration of gravity $g=1$. In these units, the box has width 48, height 460 and depth 1.2.  The unit of simulation time is equivalent to the time it takes for a sphere to fall through its own radius, and the simulation timestep in these units is $10^{-4}$.  \par

Though the vertical walls of the rectangular prism are set using the $\text{set wall/gran}$ command in LAMMPS, the base of the hopper is constructed using highly overlapping particles that are frozen in place to obtain a smooth surface. The Hertzian interaction governs interparticle and particle-wall collisions on all surfaces, and is given by the following equation \cite{Silbert2001Granular}: 

\begin{multline}
        \vec{F_\text{Hz}} = \sqrt{\delta} \sqrt{\frac{R_i R_j}{R_i + R_j}} [(k_n \delta \vec{n_{ij}} - m_\text{eff} \gamma_n \vec{v_n}) \\
        - (k_t \Delta \vec{s_t} - m_\text{eff} \gamma_t \vec{v_t})] = \vec{F}_n + \vec{F}_t
\end{multline}

The first and second terms give the normal and tangential components of the force $\vec{F}_n$ and $\vec{F}_t$, which are normal and tangential relative to the intersecting plane of the particles. Here $\delta$ is the overlap between contacting spheres, with $\vec{n_{ij}}$ the vector connecting their centers; $R_i$ and $R_j$ are the radii of the two particles (here equal), $\Delta \vec{s_t}$ is the tangential displacement vector between the particles, and $v_n$ and $v_t$ are the normal and tangential components of their relative velocity.  For our simulation, the values for normal and tangential elasticity constants are: $k_n = 200000, k_t = \frac{2}{7} k_n$; the normal damping constant is $\gamma_n = 50.0$, while the tangential $\gamma_t = 0.0$. Note that the spring constants are smaller than the corresponding material parameters for steel, but choosing larger values would make the simulation impossibly slow~\cite{Silbert2001Granular}. The tangential force $F_T$ is truncated according to the Coulomb frictional yield criterion: $F_T / F_N \leq \mu$, where we set the inter-particle friction to $\mu = 0.86$. The particle-wall friction assumes the following values in the simulation: $\mu_{W} = 0.0, 0.3, 0.6, $ and $0.9$. \par
    
To initialize each simulation run, we fill the entire container with a random arrangement of 45,000 spheres. Many of these overlap, and after deleting all overlapping pairs, we allow the spheres to fall under gravity and settle onto a plane lining the hopper bottom that prevents them from flowing through the outlet. This process leaves us with approximately $10^4$ particles up to a height of approximately 205. We wait long enough to ensure that the average kinetic energy of the particles falls below a threshold value of $ 10^{-12}$. Next, we remove the plane and allow particles to flow out under gravity. The flow rate is found by measuring the rate of change in the number of particles in the box. As in the experiment, we use 9 opening sizes (from $6d$ to $22d$ in steps of $2d$) at each $\mu_{W}$. The simulation is repeated 10 times with different initializations for each wall friction and opening size. \par
    
As in Fig.~\ref{ExperimentFlows}, we calculate the fill percentage as the ratio of number of particles remaining in the hopper to the number of particles before flow started. This is shown as a function of time in Fig.~\ref{OpeningFlows} for a single opening, $14d$, at the four different wall frictions. All four curves begin linear, and flatten as the hopper empties. The dotted lines show straight line fits to the linear region. Note that while the curves for non-zero friction are linear, there is a small deviation from linearity for $\mu_W=0$. \par

This zero wall friction case is shown for all openings in Fig.~\ref{Friction00Flows}. Here, the deviations from linearity become more evident as the opening size increases. The difference between zero and non-zero $\mu_{W}$ cases is more clearly seen in the next figure, Fig.~\ref{DerivativeFlow}, which shows the variation with time of the discrete time derivative of the number of particles in the hopper for all $\mu_{W}$ and openings. Here the $\mu_{W} = 0$ case, Fig.~\ref{DerivativeFlow}(a), clearly distinguishes itself from the others, but in a non-obvious way. For all non-zero frictional coefficients, Fig.~\ref{DerivativeFlow}(b) - (d), there is a flat region at early times of constant flow rate; the duration of that constant flow is naturally smaller for large openings, since the hopper drains faster. For zero friction, however, though the flow rate at large opening sizes decreases in magnitude over time till it vanishes, it is close to constant for smaller openings (6$d$ to 10$d$), which is a somewhat surprising result, since it is usually assumed that wall friction is essential to the constant flow rate exhibited by draining grains. This result indicates that factors such as geometric constraints become significant as the opening narrows.  \par

\section{Results}
\label{results}
 
\begin{figure*}
\begin{center}
  \includegraphics[width=1.0\textwidth]{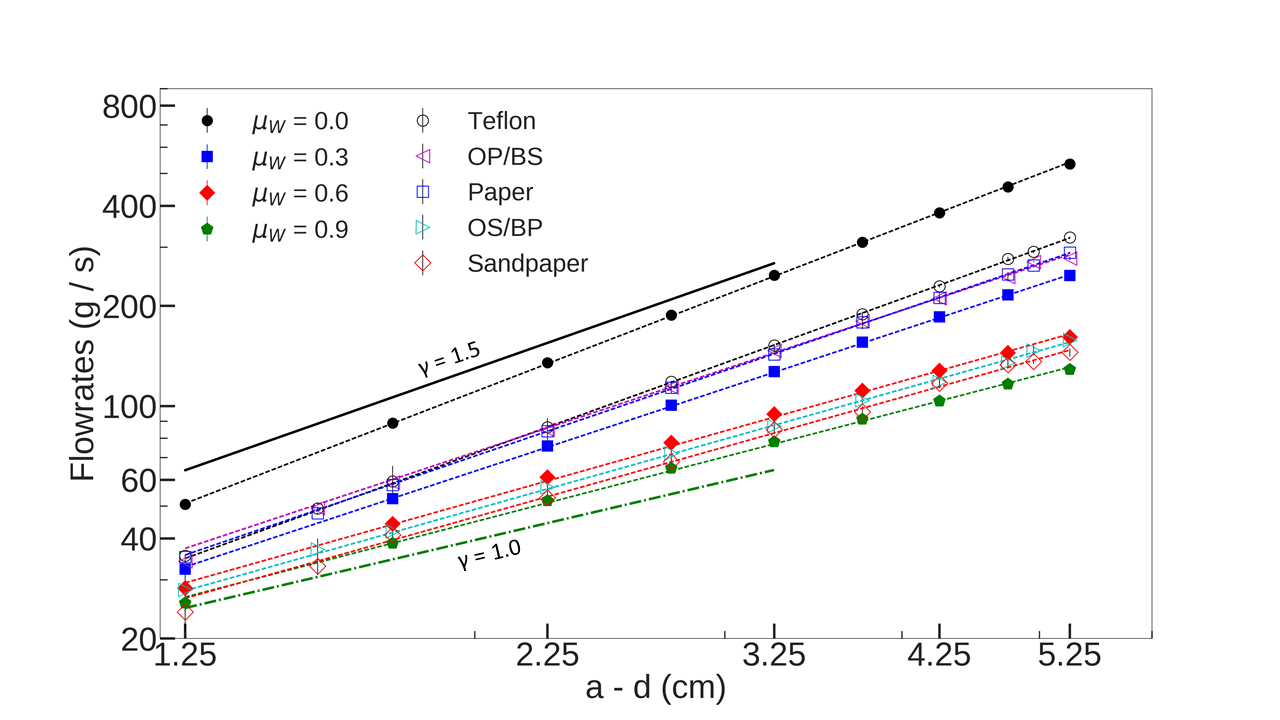}

\caption{Mass Flow rates in g/s as a function of reduced opening size for both the experiment and simulation data on a log-log scale. The simulation data is given by the solid symbols, and the experimental data by the open symbols, and both have error bars obtained by averaging over distinct runs. The simulation errors are smaller than the marker sizes. OS/BP refers to opening sandpaper and bulk paper (see left panel of Fig.~\ref{Schematic}) and OP/BS to opening paper and bulk sandpaper. The OP/BS and paper data are congruent to one another, as are the OS/BP and sandpaper data. The dotted lines are power law fits to the data, where the exponents of power law fit along with their standard deviation are given in Table \ref{Tab:Stokes}. Other than the $\mu_W=0$ data from simulation, the slopes fall between the solid black line of power law exponent 1.5 and the green dash-dot line of exponent 1.0.}
\label{FlowvsOpening}
\end{center}
\end{figure*}

We now look at how the flow rate for both experiment and simulation varies with opening size, and how this dependence is affected by wall friction. The flow rates are extracted from linear fits of the data for the variation in time of the number of particles in the hopper, some of which have been shown in the previous figures. These mass flow rates, converted to units of g/s, are plotted vs. reduced opening size $(a-d)$ in Fig.~\ref{FlowvsOpening}: experimental data (open symbols), each averaged over 5 runs for each distinct wall lining; and simulation data (solid symbols), each point averaged over the 10 runs at each of the four friction values. Note that the experimental OS/BP data, corresponding to the experimental hopper lining where the bulk is lined with paper and the opening with sandpaper (see the left panel of Fig.~\ref{Schematic}), lie on top of the data for the fully sandpaper-lined hopper, seen in the right panel of Fig.~\ref{Schematic}. Likewise the OP/BS data, which has paper at the opening and sandpaper in the bulk, is indistinguishable from the purely paper-lined hopper data. This tells us the surprising result that the mass flow rate is primarily affected by the value of the wall friction in the vicinity of the opening, and not in the bulk. \par

The values of the particle-wall static friction coefficient used in simulation were chosen to resemble those of the experimental materials being used, and indeed, we observe in Fig.~\ref{FlowvsOpening} that the simulation data for each wall friction follows the corresponding data in the experiment. The error bars for each data point are small enough that most of them are contained within the symbol size. The dashed lines represent power law fits to the data on this log-log graph, and these are clearly not parallel to one another. Though the dynamical range in opening sizes is necessarily small (the hopper clogs intermittently at smaller openings, and at larger openings, empties too rapidly to establish a steady rate), the fits indicate that the exponent of the power-law dependence of flow rate on reduced opening size decreases as wall friction increases. The power law exponents of all the experimental and simulation data, shown in Table \ref{Tab:Stokes}, lie between 1.5 (black solid line in Fig.~\ref{FlowvsOpening}) and 1.0 (green dot-dashed line), and the difference from one exponent to the next is larger than the standard deviation in each. The exception is the zero friction $\mu_{W}=0$ case, which has an exponent that appears slightly higher than 1.5, and we think this is because the hopper drains at a faster rate than linear at early times for large openings. \par

\begin{figure*}
\begin{center}
  \includegraphics[width=1.0\textwidth]{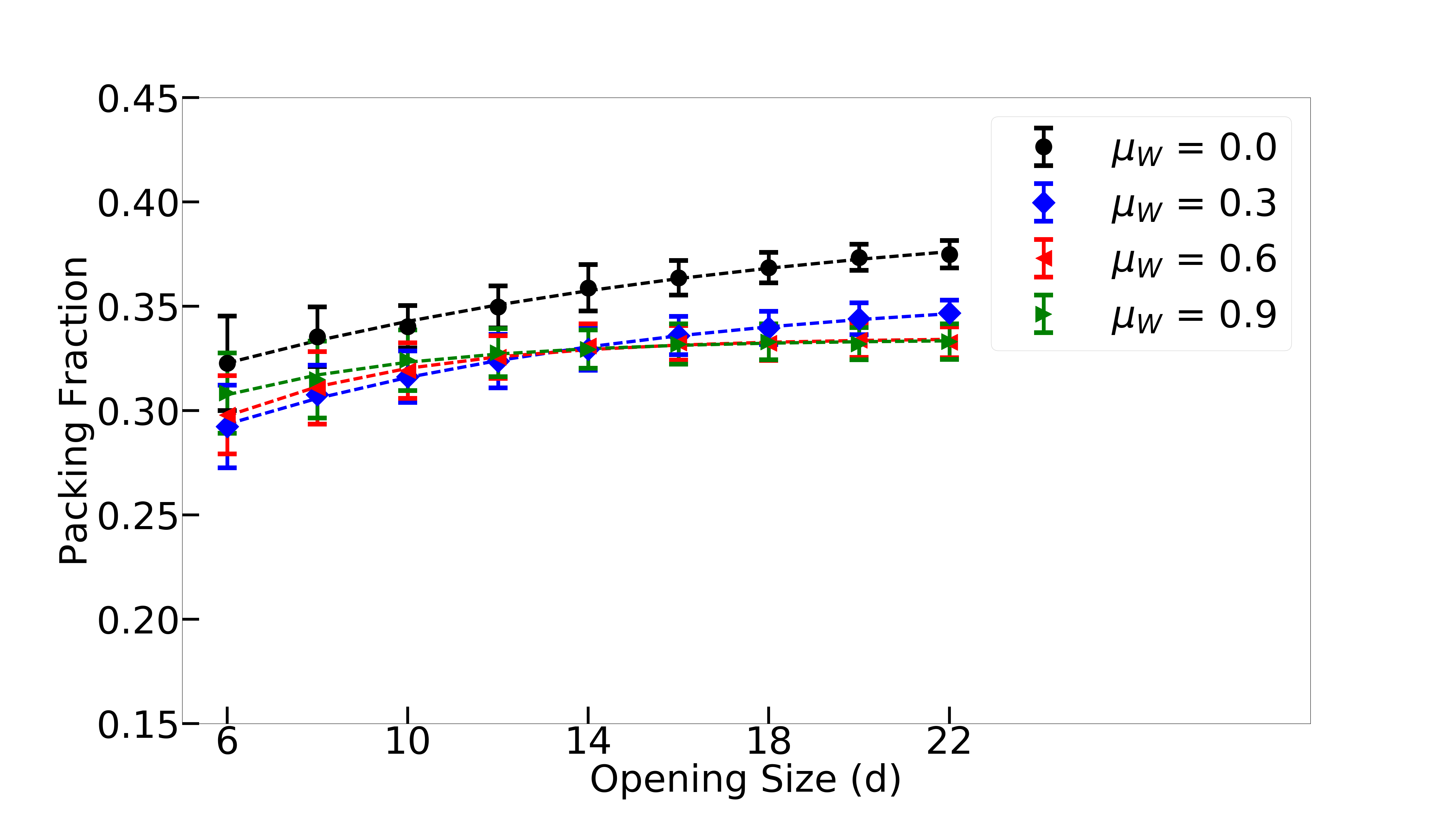}
\caption{Packing fractions within a box of width equal to the opening size, height $1d$ and depth $1.2d$ at the hopper opening as a function of the opening size. The values plotted represent the average volume fraction in the box over the duration of the flow with vertical bars showing one standard deviation. The dashed lines represent an empirical fit of the form: $B(1 - C e^{-a/L})$ \cite{Janda2012Flow}, where $B$, $C$ and $L$ are fitting parameters. In our fits, $B$ and $C$ do not vary much, while $L$ changes from about 13 for $\mu_W = 0$ to 4.5 for $\mu_W = 0.9$.}
\label{PackingFraction} 
\end{center}
\end{figure*}

The mass flow rate at the hopper exit is a function of two quantities: the packing fraction and the velocity of the flowing particles at the opening. We examine both separately in the simulations to glean where wall friction plays the bigger role. In Fig.~\ref{PackingFraction}, we plot the mean volume packing fraction, measured just above the outlet, as a function of opening size, for all 4 wall friction values. To do so, we construct a box centered at the opening of height 1$d$, depth $1.2d$, and width equal to the opening size, and find the time-average and standard deviation of the packing fraction measured from snapshots of the simulation taken every $10^4$ timesteps. While the mean packing fraction is higher when wall friction is zero, we see no significant difference in the curves for all non-zero $\mu_{W}$ values. For a given $\mu_W$, the packing fraction variation with opening size is relatively slight as compared to the accompanying variation of particle velocities. The largest change in packing fraction, around 10\%  from $a = 6$ to $a = 22$, occurs for $\mu_W = 0$, with the corresponding exit velocity changing by 300\%, see Fig.~\ref{VelocityDistributions}.  The variation of packing fraction with opening size is even less significant when we do a similar analysis within the bulk at a height greater than $10d$ (not shown here). We have also checked that dividing the mass flow rate by the corresponding packing fraction cannot account for the apparent change in the Beverloo exponent shown in Fig.~\ref{FlowvsOpening} and Table~\ref{Tab:Stokes}. \par

We find this variation of packing fraction with opening size near the outlet is well fit by an empirical model from Janda et al. \cite{Janda2012Flow}. This fit, shown by the dashed lines in Fig.~\ref{PackingFraction}, has the form $B(1 - C e^{- a/L})$, where we find that the parameter $L$ is the most sensitive to friction. This particular form \cite{Janda2012Flow} was motivated by experimental results from Mankoc et al. \cite{Mankoc2007Flow}, who showed that Beverloo's law tends to fail for small openings, and $L \approx 6$ represents an opening size below which the likelihood of clogging is high and the flow rate is suppressed due to changes in the density at the outlet. Our fits show that this parameter $L$ decreases as wall friction increases, subtly altering the shape of the packing fraction dependence on opening size. This also suggests, somewhat counter-intuitively, that the length scale over which the clogging affects the packing fraction at the exit is smaller at higher friction. \par

Next, we focus on the effect of changing friction and opening size on particle velocities at the exit. We analyze these in a few different ways. First, we look into the putative notion of a {\it free-fall arch}. For a particular wall friction and opening size, we find instantaneous velocities for spheres at the exit from snapshots of the flow taken at every simulation time. These are then converted to a height $h$ by setting $h = v^2/2g$, which can be interpreted as the starting height for a particle had it been in free-fall. The dots in Fig.~\ref{SampleVelocityDistribution} show the resulting heights for each particle (one dot per particle for each time) as a function of its x-position, where the color represents the magnitude of the velocity ranging from low (dark blue) to high (yellow). We then divide the x-axis into bins of 0.25$d$ width and average the $v^2/2g$ values in each bin, giving the red solid line running through the data, where the vertical bars represent the standard deviation in each bin. \par

The averaged height profiles with error bars for all the openings studied are shown in Fig.~\ref{VelocityDistributions} for each opening size, with each panel representing a different $\mu_{W}$. At zero particle-wall friction, panel (a), $\langle v^2/2g \rangle$ at the exit scales with opening size, consistent with the expectation of a free fall arch. This is made clearer in Fig.~\ref{ScaledVelocityDistribution}(a), which shows the same figure with {\it both} x and y axes scaled by the opening size. The curves for all openings collapse within error bars, and this shape is well fit by a parabolic function, shown by the solid black line in the figure, in good agreement with the shape seen for the velocity profiles in experiments\cite{Janda2012Flow}. This suggests that the description of a free fall arch does not seem unreasonable in the $\mu_W = 0$ case. \par

However, when we examine this further by looking at spatial distributions of the actual particle velocities, we encounter an apparent contradiction. To visualize this, we coarse-grain the hopper into square cells of side $1d$ and find the time-averaged sphere velocity vector in each cell.
The top row of Fig.~\ref{VelocityVectors} shows, for $\mu_W = 0$, the time-averaged velocity at spatial points in the lower half of the hopper. The figures on the left and the right represent two opening sizes. Note that the coarse-grained velocity vectors do not begin to increase until the very bottom of the hopper, implying that the acceleration region for $\mu_W=0$ is far smaller than the opening size, and there is no real free-fall arch. We conclude that it is misleading to extract anything definitive about the nature of the acceleration experienced by particles simply from observing their velocities at the exit. \par 

The picture changes for non-zero wall friction. The remaining panels in Fig.~\ref{VelocityDistributions} show that the $\langle v^2/2g \rangle$ distributions do not scale as expected with the opening size for all finite values of friction. The vertical axis scale for Fig.~\ref{VelocityDistributions}(b)-(d) have a maximum of 12, in contrast to the vertical axis of Fig.~\ref{VelocityDistributions}(a) for $\mu_W=0$, which has a maximum of 36. This already shows that as $\mu_W$ increases, the mean square velocity at the exit, or free-fall arch height, scales more weakly with opening size. At $\mu_W= 0.9$, Fig.~\ref{VelocityDistributions}(d), the average velocity at the opening becomes fully independent of opening size. The shape of the arch also changes, and is no longer parabolic. \par

The change in shape with friction is more clearly seen in Fig.~\ref{ScaledVelocityDistribution}(b), where all the curves collapse when only the x-axis is scaled by the opening $a$. This shape is no longer parabolic, but closer to the form $A(1 - B\exp[-C(|x/a| - 1)])$, where $A$ is set by the maximum value of $\langle v^2/2g \rangle$ for each opening, and $B, C$ are fitting parameters. This suggests that even though the na\"{i}ve notion of a free-fall arch is no longer valid for large values of wall friction, the opening size remains the only relevant length scale in the problem.   \par

To better understand the origin of the change in exit velocity distributions with wall friction, we return to the analysis of the spatial distribution of coarse-grained velocity vector fields in the hopper. Deviations from mean flow velocity in the lower half of the hopper are shown in Fig.~\ref{VelocityVectors}, where left and right represent opening sizes $10d$ and $22d$ respectively, and each row represents a different wall friction $\mu_W$.
The color bar indicates the magnitude of the coarse-grained velocity vector. We observe from the change in vector size that there is a transition from bulk flow to rapid flow near the opening where the velocity is changing. To further accentuate this transition, the curves in blue (dashed)/black (solid) represent the height at which the velocity increases 10\%/50\% over the mean particle velocity in the bulk. Thus these curves also indicate a spatial boundary from a region of constant mean particle velocity to a region of accelerating flow.  \par

Whether you look at the velocity vector magnitudes using the color bar, or the dashed blue/solid black curves, there is a clear difference in shape between the zero and non-zero $\mu_W$ values. For the $\mu_W=0$ case, top row, particles flow on average at the same rate till they get to a region less than $20d$ in height, whereupon they accelerate rapidly out of the opening. This is also seen in movies of the simulation, where the particles appear to be unaware of the exit almost until they reach the bottom of the hopper. After reaching the bottom, particles flow laterally towards the exit. This lateral flow is seen more clearly for the larger opening: top row, right. The size of the acceleration region does not change significantly with opening size ($14d$ on the left, $22d$ on the right). However, particle velocities as they exit do increase with the size of the opening. \par

In the non-zero friction cases, a conical region of the hopper where particles are moving is visible, within which the transition from steady to accelerated flow is demarcated by a change in color for all $\mu_W\neq 0$, rows 2 - 4 of Fig.~\ref{VelocityVectors}. (Note that the color bar for rows 2 - 4 is the same, i.e. for all non-zero friction, but half the size of the color bar for the zero friction case, top row). The width of this conical region changes with opening size, but for a given opening, there is not much change from one wall friction to the next. At all $\mu_W$, comparing the height at which the acceleration starts, we see not much difference from opening $14d$ on the left to opening $22d$ on the right. What does change as friction increases is {\it how much acceleration occurs}, judging by the change in size of the vectors from the start of the acceleration region to the exit. While the exit velocities increase significantly with opening size for $\mu_W = 0.3$, this is negligible at $\mu_W=0.9$ - this is consistent with the lack of change in the height of the exit velocity profiles with opening size seen for the same wall friction, Fig.~\ref{VelocityDistributions}(d). \par

The decrease in the magnitude of vertical acceleration with increasing friction $\mu_W$ is highlighted in Fig.~\ref{Accelerations}. Here, we calculate the vertical velocity component $v_y(t)$ in a series of square cells of size $1d$ running down the center of the hopper, and use it to calculate $a_y= dv_y/dt = v_y dv_y/dy$ and plot it as a function of vertical position $y$. This gives us the rather surprising result that the average acceleration at the hopper center for $\mu_W=0$ becomes larger than $g$, close to $6g$ at the exit. This occurs even when the size of the averaging box is increased. We believe this is due to the fact that the flow remains collisional up until the hopper exit - the instantaneous accelerations come from collisional impulses from other particles and from the front and back walls - and these can be very large in the absence of friction. As $\mu_W$ increases, the magnitude of the acceleration drops, but the accelerations at all $\mu_W \neq 0$ begin to deviate from zero at about the same height for a given opening, consistent with what was seen in Fig.~\ref{VelocityVectors} -- the deviation occurs at a smaller value of $y$ for $\mu_W = 0$. Notably, far from taking a constant free-fall value, accelerations continue to increase as the particles approach the opening, a pattern that persists for all openings and all values of $\mu_W$. \par

\section{Discussion}
\label{discussion}

\begin{figure*}
\begin{center}
  \includegraphics[width=1.0\textwidth]{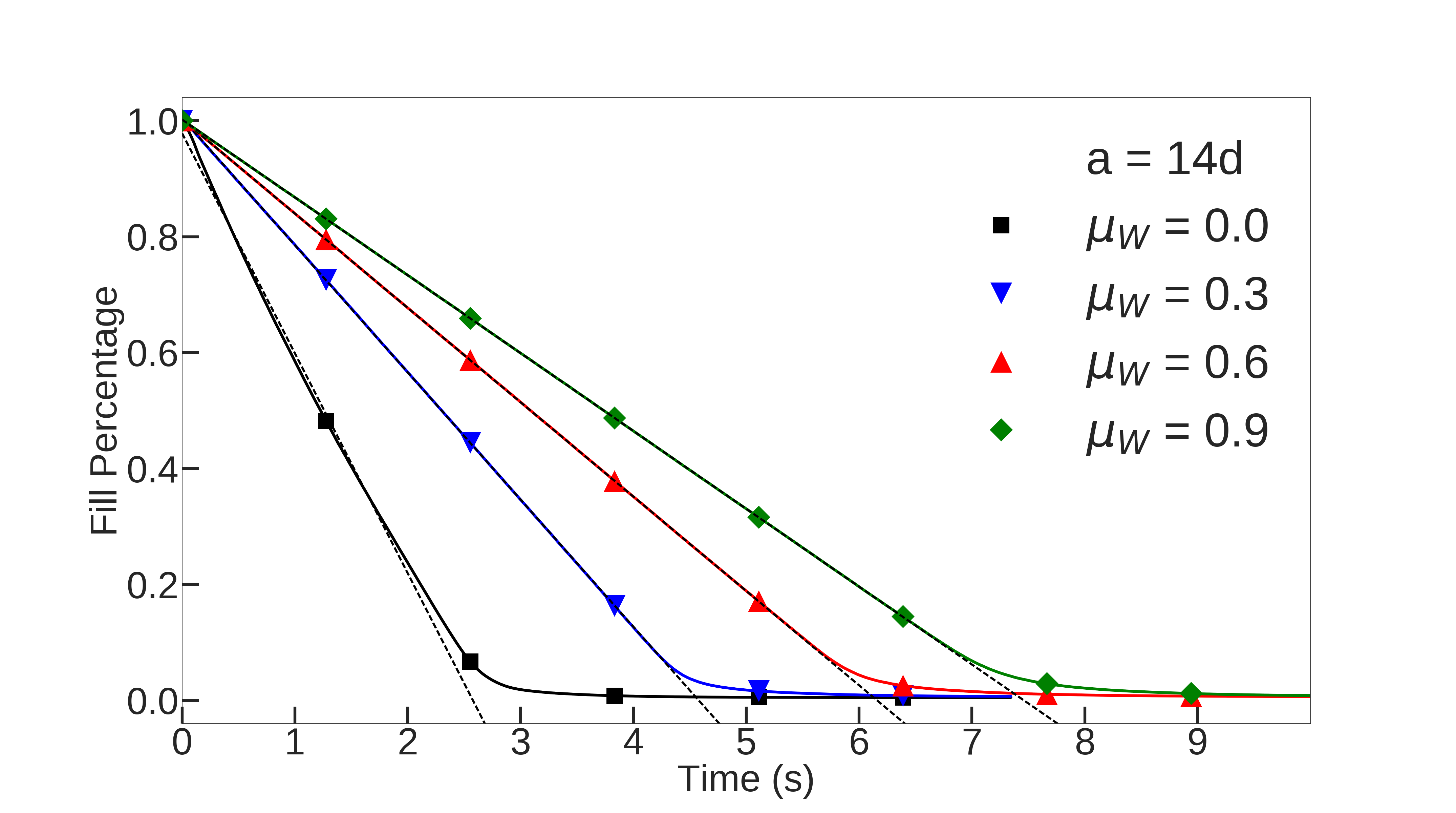}
\caption{Fill percentage vs. time for simulations at fixed opening size $a$ = 14$d$ and varying wall frictions. The markers are not data points, but are instead used to identify the curve they are on. The lines shown here are actually connected dots, and there are approximately 7000 data points for each opening size, with the markers placed every 500 data points. The dashed lines represent linear fits that give us the flow rate for each case. There is deviation from linearity when the hopper empties for all wall frictions, but the $\mu_W = 0.0$ case shows deviations from linearity during flow as well.}
\label{OpeningFlows}
\end{center}
\end{figure*}

We have examined the effect of changing particle-wall friction on the flow patterns of spheres falling under gravity in a quasi-2D hopper of depth 1.2$d$. What appears to matter are particle collisions with the front and back walls, since changing the friction at the side walls has no effect. We find that the flow rate is constant at all non-zero wall friction $\mu_W$ and opening sizes, Fig.~\ref{OpeningFlows}. While the rate of flow in the absence of wall friction, $\mu_W=0$, is not constant, as seen in Figs.~\ref{Friction00Flows} and \ref{DerivativeFlow}(a), the flow behavior is distinctly different from Newtonian fluid flow, in which the flow rate would decrease linearly in time. The mass flow rate also approaches a constant value for small openings, Fig.~\ref{DerivativeFlow}(a), which leads us to conclude that wall friction is not the only requirement to have a constant rate of flow - geometric confinement appears to also play an important role. \par

While the dependence of the flow rate {\it magnitude} on the particle-wall friction coefficient $\mu_W$ comes as no surprise, Fig.~\ref{OpeningFlows}, it is less obvious that the flow rate scaling with opening size changes as a function of friction, and no longer follows the expected Beverloo scaling exponent of $3/2$, Fig.~\ref{FlowvsOpening}. Since the dependence of packing fraction on opening size in the regime we have studied is weak, Fig.~\ref{PackingFraction}, this decrease in the exponent comes primarily from how the particle velocity field transitions from bulk flow to accelerated flow. As seen in Fig.~\ref{VelocityVectors}, the region where the acceleration begins stays roughly the same for all values of wall friction $\mu_W$, but the primary effect of friction is to impede the acceleration of the particles. Fig.~\ref{Accelerations} shows that the magnitude of this acceleration is strongly affected by friction. Indeed, the high values of these accelerations at low friction, which come from taking finite differences of particle positions separated by 0.1 simulation times, lead us to believe the particles are not, in fact, in free fall, but are undergoing many collisions right until they exit. \par

Thus, in conclusion, our experiments and simulations indicate that the idea of a free-fall arch should not be interpreted literally, but as a region that marks the boundary between steady flow and accelerated flow, a region where the dynamics of particle-particle and particle-wall collisions begin to play a role in how rapidly grains exit the hopper. The height of this region is of order the opening size, consistent with the fact that the ratio of opening size to particle diameter is the only relevant dimensionless scale in the problem. This is borne out by the experiments we reported on here, which showed no statistical difference between the flow rate of a hopper where the entire wall was covered with a particular frictional backing, and one in which merely the region of height 22$d$ near the exit of the hopper had that frictional backing. Note that even when we set particle-wall friction to zero in our simulations, inter-particle friction remains non-zero. The fact that the dependence on particle-wall friction of the flow rate scaling with opening size vanishes in our 3D experiments and simulations is consistent with previous experiments that have shown that while flow in 2D experiments is highly collisional, 3D flow tends to be less collisional and have more enduring frictional contacts \cite{3Dflowexpt}. It remains to be investigated at what hopper thickness this transition from 2D to 3D flow occurs. Finally, we plan in the future to measure a dynamical stress tensor during flow to see if the Janssen effect continues to hold during flow, and whether any change is seen as a function of changing particle-wall friction. \par

\section{Acknowledgements}
\label{acknowledgements}

    We thank Narayanan Menon for helpful discussions, and Mollie Pleau for sharing her 3D data. SKB acknowledges funding from ICTS Bangalore, the APS-IUSSTF travel grant, and NSF DMR 1506750.

\bibliographystyle{spphys}       
\bibliography{JournalArticle.bib}   

\begin{thebibliography}{10}
\providecommand{\url}[1]{{#1}}
\providecommand{\urlprefix}{URL }
\expandafter\ifx\csname urlstyle\endcsname\relax
  \providecommand{\doi}[1]{DOI \discretionary{}{}{}#1}\else
  \providecommand{\doi}{DOI \discretionary{}{}{}\begingroup
  \urlstyle{rm}\Url}\fi

\bibitem{Nedderman1982Flow}
R.~Nedderman, U.~T{\"u}z{\"u}n, S.~Savage, G.~Houlsby, Chem. Eng. Sci
  \textbf{37}(11), 1597 (1982)

\bibitem{Beverloo1961Flow}
W.A. Beverloo, H.A. Leniger, J.~Van~de Velde, Chemical engineering science
  \textbf{15}(3-4), 260 (1961)

\bibitem{Tighe2007Pressure}
B.P. Tighe, M.~Sperl, Granular Matter \textbf{9}(3-4), 141 (2007)

\bibitem{Hilton2011Granular}
J.~Hilton, P.~Cleary, Physical Review E \textbf{84}(1), 011307 (2011)

\bibitem{Vivanco2012Dynamical}
F.~Vivanco, S.~Rica, F.~Melo, Granular Matter \textbf{14}(5), 563 (2012)

\bibitem{Rubio2015Disentangling}
S.M. Rubio-Largo, A.~Janda, D.~Maza, I.~Zuriguel, R.~Hidalgo, Physical Review
  Letters \textbf{114}, 238002 (2015)

\bibitem{NeddermanBook1992}
R.M. Nedderman, \emph{Statics and Kinematics of Granular Materials} (Cambridge
  University Press, 1992).
\newblock \doi{10.1017/CBO9780511600043}

\bibitem{Smith}
B.~Carballo-Ramirez, M.~Pleau, N.~Easwar, S.~Birwa, N.~Shah, S.~Tewari.
\newblock Investigation of the effect of wall friction on the flow rate in 2d
  and 3d granular flow.
\newblock \url{http://meetings.aps.org/link/BAPS.2016.MAR.P43.10} (2016)

\bibitem{Janssen1895}
H.~Janssen, Z. Ver. Dtsch. Ing. \textbf{39}(35), 1045 (1895)

\bibitem{Bertho2003Dynamical}
Y.~Bertho, F.~Giorgiutti-Dauphin{\'e}, J.P. Hulin, Physical review letters
  \textbf{90}(14), 144301 (2003)

\bibitem{Aguirre2010}
M.A. Aguirre, J.G. Grande, A.~Calvo, L.A. Pugnaloni, J.C. G{\'e}minard,
  Physical Review Letters \textbf{104}(23), 238002 (2010)

\bibitem{LAMMPS}
Lammps molecular dynamics simulator.
\newblock \url{https://lammps.sandia.gov/index.html}.
\newblock Accessed: 2016-2019

\bibitem{Silbert2001Granular}
L.E. Silbert, D.~Erta{\c{s}}, G.S. Grest, T.C. Halsey, D.~Levine, S.J.
  Plimpton, Physical Review E \textbf{64}(5), 051302 (2001)

\bibitem{Janda2012Flow}
A.~Janda, I.~Zuriguel, D.~Maza, Physical review letters \textbf{108}(24),
  248001 (2012)

\bibitem{Mankoc2007Flow}
C.~Mankoc, A.~Janda, R.~Arevalo, J.~Pastor, I.~Zuriguel, A.~Garcimart{\'\i}n,
  D.~Maza, Granular Matter \textbf{9}(6), 407 (2007)

\bibitem{3Dflowexpt}
E.~Gardel, E.~Seitaridou, K.~Facto, E.~Keene, K.~Hattam, N.~Easwar, N.~Menon,
  Philosophical Transactions of the Royal Society A: Mathematical, Physical and
  Engineering Sciences \textbf{367}, 5109 (2009)

\end{thebibliography}
\nocite{*}


\newpage

\begin{table}[ht]
\begin{center}
\begin{tabular}{| c | c |}
\hline

 $\mu_W$ & $\gamma$ \vspace{0.0cm}\\\hline
 $0.0$ & $1.65 \pm 0.01$ \vspace{0.0cm}\\\hline 
 $0.3$ & $1.41 \pm 0.01$ \vspace{0.0cm}\\\hline
 $0.6$ & $1.20 \pm 0.02$ \vspace{0.0cm}\\\hline
 $0.9$ & $1.11 \pm 0.01$ \vspace{0.0cm}\\\hline
 Teflon & $1.55 \pm 0.01$ \vspace{0.0cm}\\\hline
 OP/BS & $1.42 \pm 0.03$ \vspace{0.0cm}\\\hline 
 Paper & $1.46 \pm 0.01$ \vspace{0.0cm}\\\hline
 OS/BP & $1.20 \pm 0.02$ \vspace{0.0cm}\\\hline
 Sandpaper & $1.20 \pm 0.02$ \vspace{0.0cm}\\\hline
\end{tabular}
\caption{Exponents for power law fits of flow rate vs. opening size shown in Fig. \ref{FlowvsOpening}. The first four rows represent the power law exponents $\gamma$ for simulation data vs wall friction coefficient, and the remaining for the experimental data vs the material of the wall backing. The uncertainty is obtained from the fit. }
 \label{Tab:Stokes}
\end{center}
\end{table}

\newpage

\newpage

\newpage

\begin{figure*}
\begin{center}
 \includegraphics[width=1.0\textwidth]{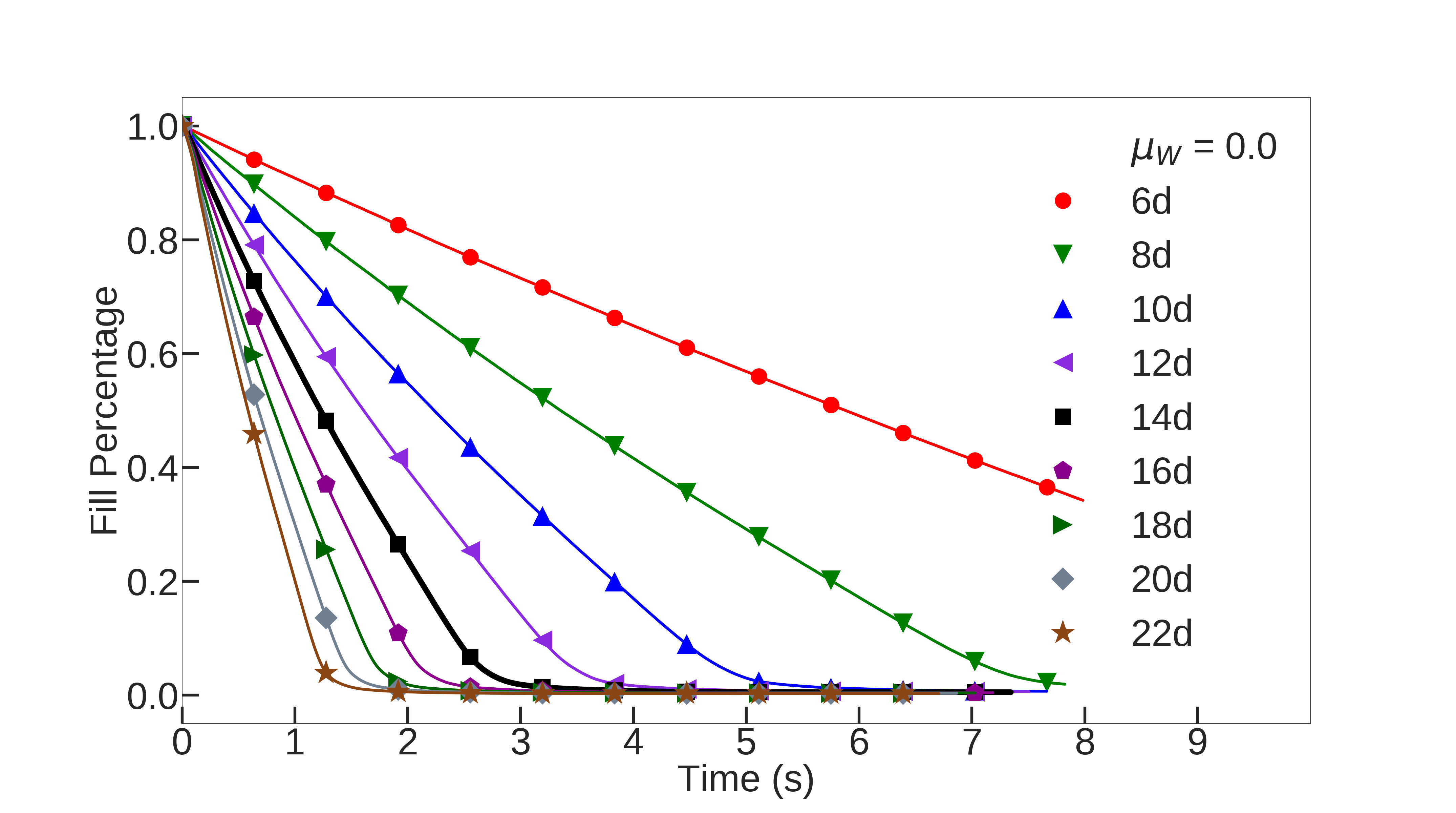}
\caption{Fill percentage vs. time at different opening sizes for fixed wall friction 0.0.  Flow rates are obtained from linear fits. The flattening of the curves represents the container running out of particles.}
\label{Friction00Flows} 
\end{center}
\end{figure*}

\newpage

\begin{figure*}
	\centering
    \includegraphics[width=1.0\textwidth]{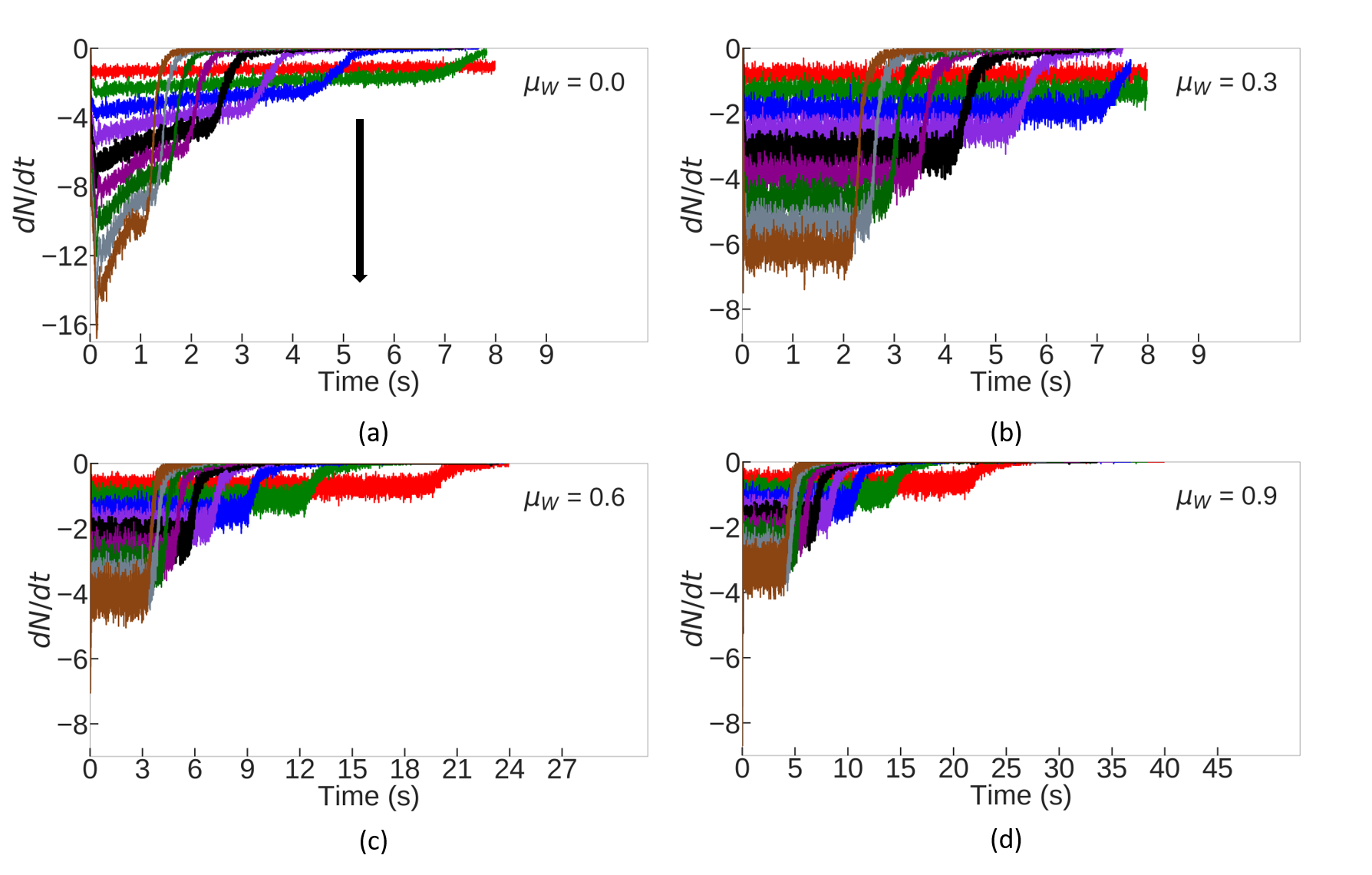}
	\setcounter{figure}{4}
	\caption{Discrete time derivative of the number of particles $N(t)$ in the hopper shown as a function of time for all openings, where panels (a) - (d) represent increasing wall friction $\mu_W$ values. Each time point $i$ here is given by the following: $\Big( \frac{dN}{dt} \Big)_i = \frac{N_{i + 1} - N_{i - 1}}{2}$. The arrow in (a) represents the direction of increasing opening size from $6d$ to $22d$ (as ordered at t = 0), for all sub-figures.  The $\mu_W = 0.0$ case shows deviation from constant flow, and these deviations increase with the opening size.}.
	\label{DerivativeFlow}
\end{figure*}

\newpage

\newpage 

\newpage

\begin{figure*}
\begin{center}
  \includegraphics[width=1.0\textwidth]{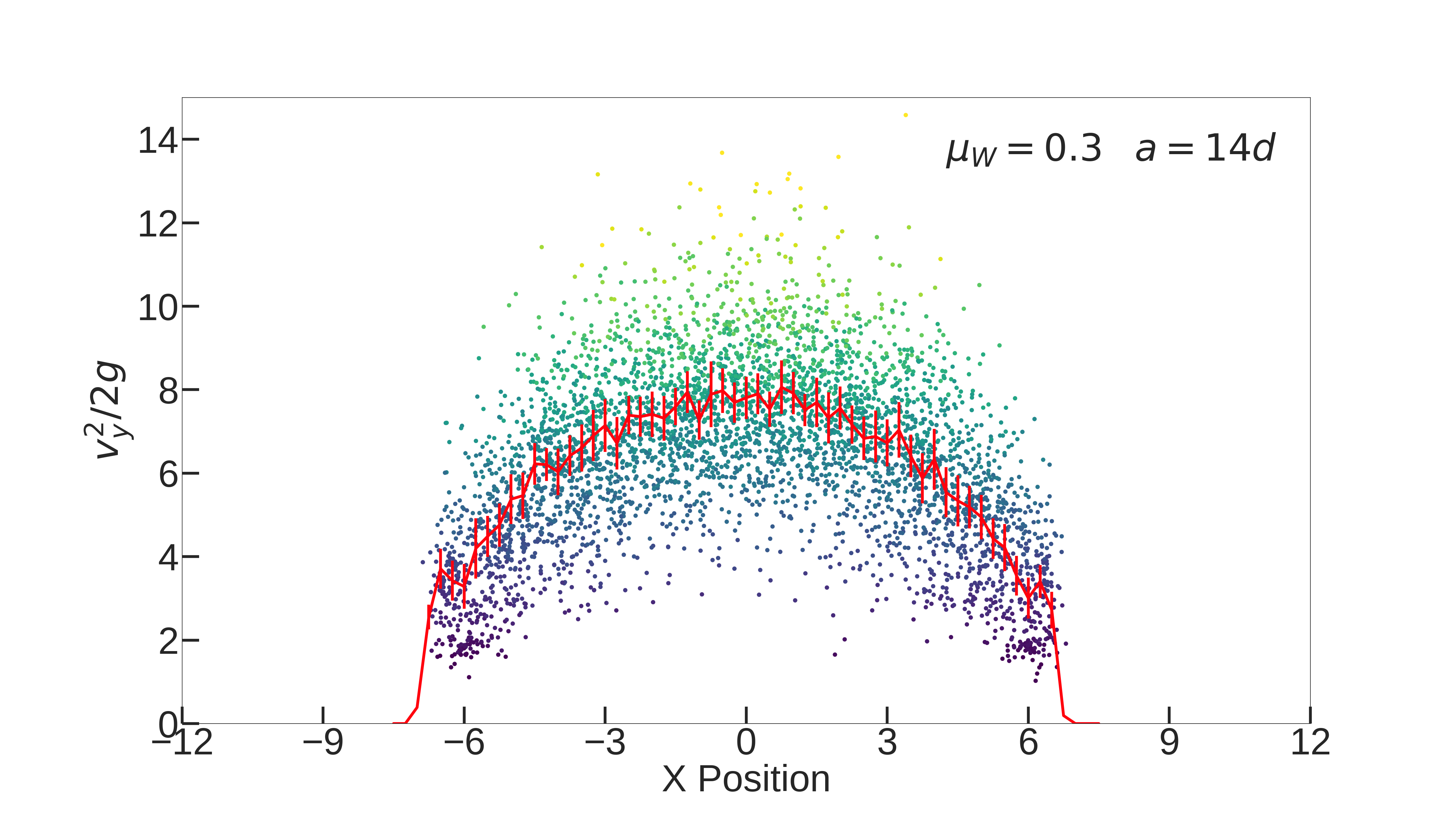}
\caption{Scatter plot of extrapolated free-fall heights, or $v^2/2g$ values, for each particle with velocity v at the opening, sampled over the entire flow period and plotted against its x-position for $\mu_W = 0.3$ and opening size $14d$. The color is representative of the "height" or velocity of the particles, with darker representing lower velocity. The corresponding distribution, obtained by averaging these points in x-bins of size $0.25d$, is shown as a solid red line, with vertical error bars representing one standard deviation.}
\label{SampleVelocityDistribution} 
\end{center}
\end{figure*}

\newpage

\begin{figure*}
	\centering
    \includegraphics[width=1.0\textwidth]{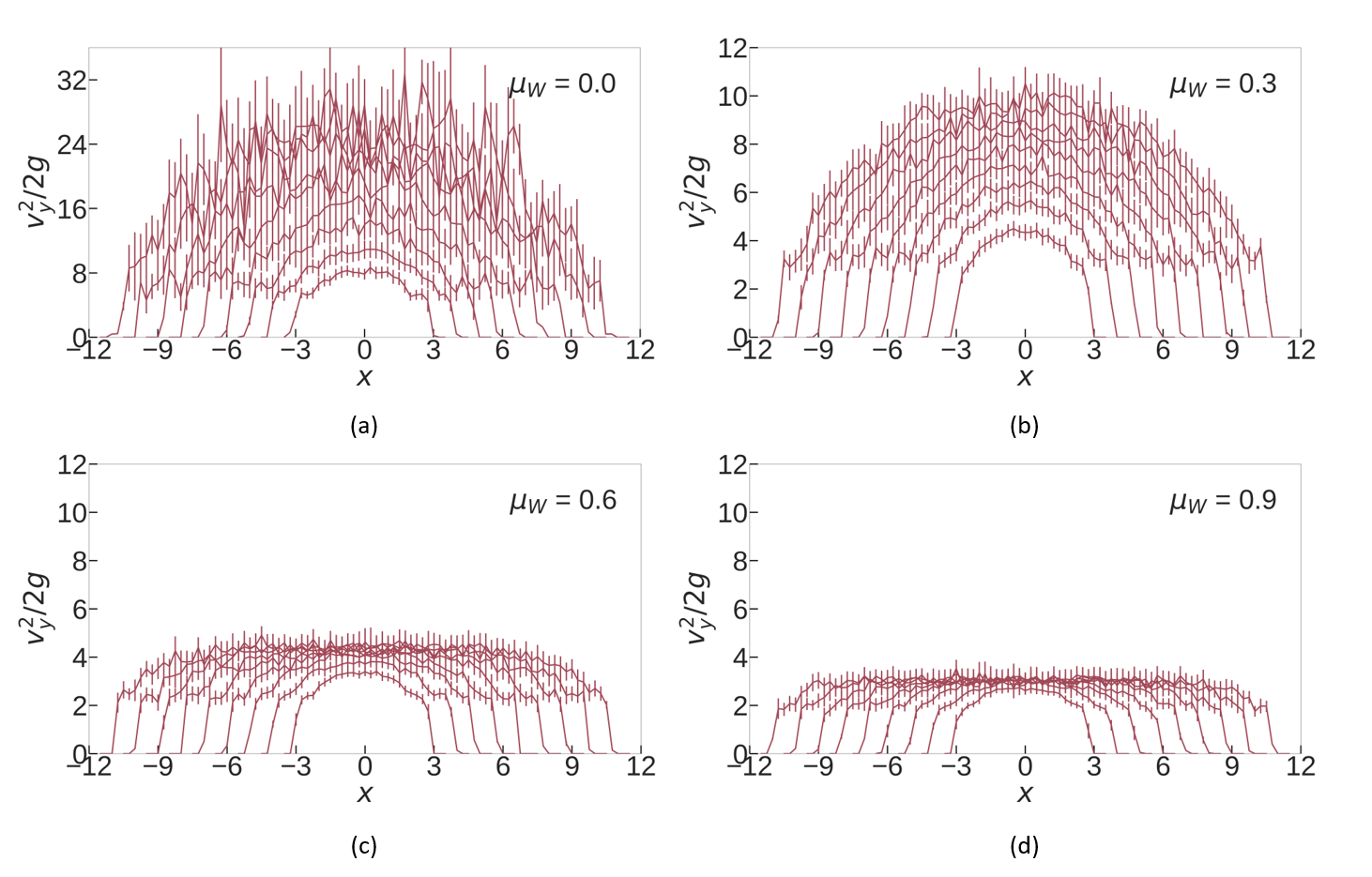}
	\setcounter{figure}{8}
	\caption{The four panels show the time-averaged profile of extrapolated `free fall heights', or $\langle v^2/2g \rangle$, along with errors bars, plotted as a function of x position for openings ranging from $6d$ to $22d$. Each panel represents a different friction $\mu_W$. Note that as wall friction increases, the `arch' heights stop increasing with the opening size.}
	\label{VelocityDistributions}
\end{figure*}

\begin{figure*}
\begin{center}
\begin{center}
  \includegraphics[width=1.0\textwidth]{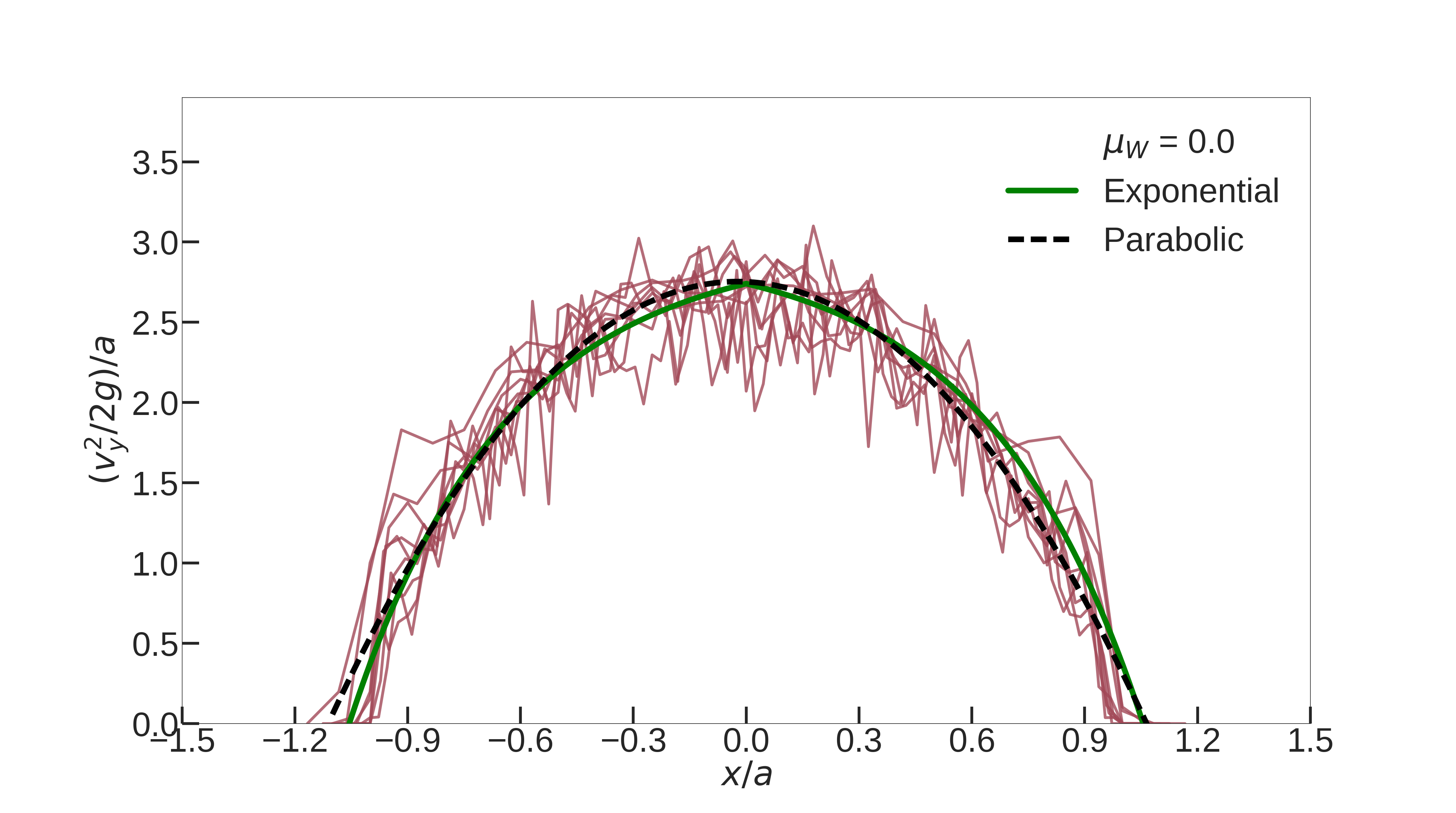}
  \captionsetup{labelformat=empty}
  \caption{$(a)$}
  \includegraphics[width=1.0\textwidth]{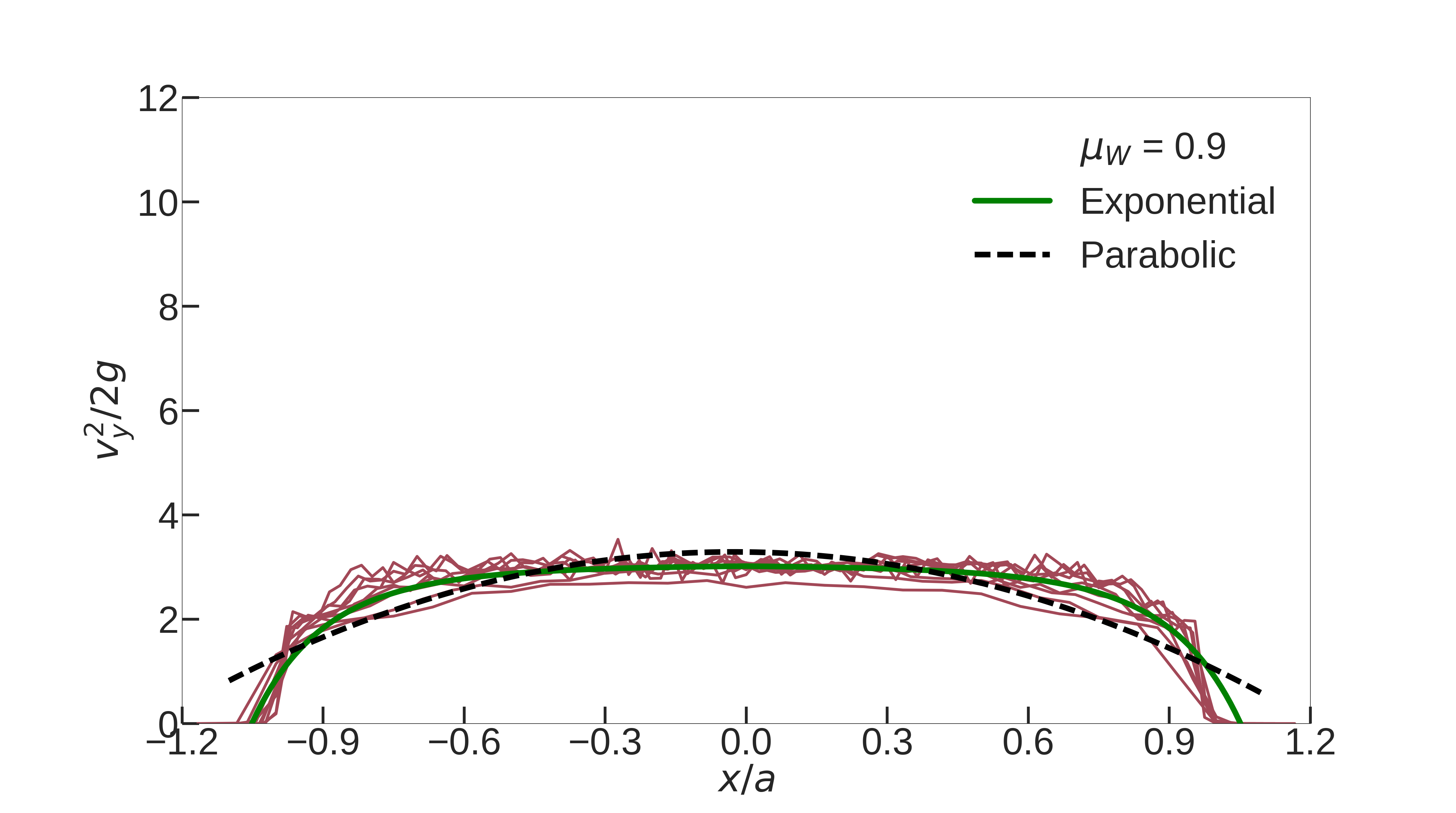}
  \captionsetup{labelformat=empty}
  \caption{$(b)$}
\end{center}
\setcounter{figure}{9}
\caption{(a) The `free-fall height' profiles $\langle v^2/2g \rangle$ for all openings at wall friction $\mu_W=0.0$ collapse when scaled along {\it both axes} by the size $a$ of the opening. The dashed black line represents a fit to a parabolic profile, and the solid green line to the form $A(1 - B\exp[-C(|x/a| - 1)])$. (b) The `free-fall height' profiles $\langle v^2/2g \rangle$ for all openings at wall friction $\mu_W=0.9$ collapse when scaled along {\it only} the x-axis by the size $a$ of the opening - the height is independent of the opening size as seen in Fig.~\ref{VelocityDistributions} (d)}
\label{ScaledVelocityDistribution} 
\end{center}
\end{figure*}

\newpage

\begin{figure*}
\begin{center}
\includegraphics[width=1.0\textwidth]{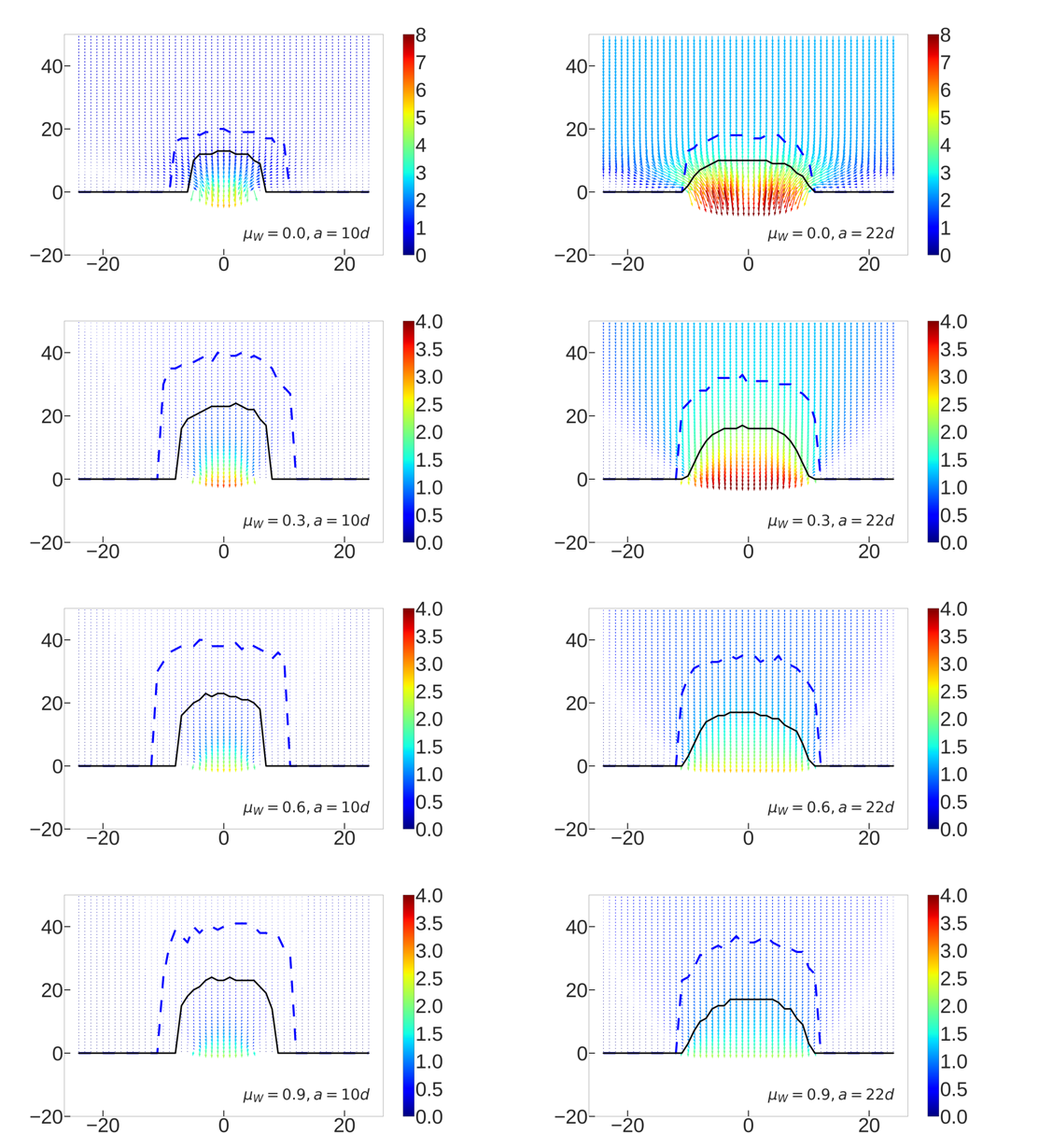}
\caption{Velocity vector field for two openings (left column $14d$, right column $22d$), where each row represents wall friction values $\mu_W = 0, 0.3, 0.6, 0.9$. The hopper is binned into square cells of side $0.25d$ to obtain this plot, where the velocity vector in each cell is the time average of particle velocities in that cell over the course of the run. The blue dashed (black solid) curve shows the points at which the average velocity of the particles becomes 10\% (50\%) larger than the constant bulk velocity. This is done to highlight the transition from bulk flow (zero acceleration) to near-opening flow (increasing acceleration). The color bar is scaled differently for $\mu_W = 0.0$ in order to capture the full range of velocities.}
\label{VelocityVectors} 
\end{center}
\end{figure*}

\newpage

\begin{figure*}
	\centering
    \includegraphics[width=1.0\textwidth]{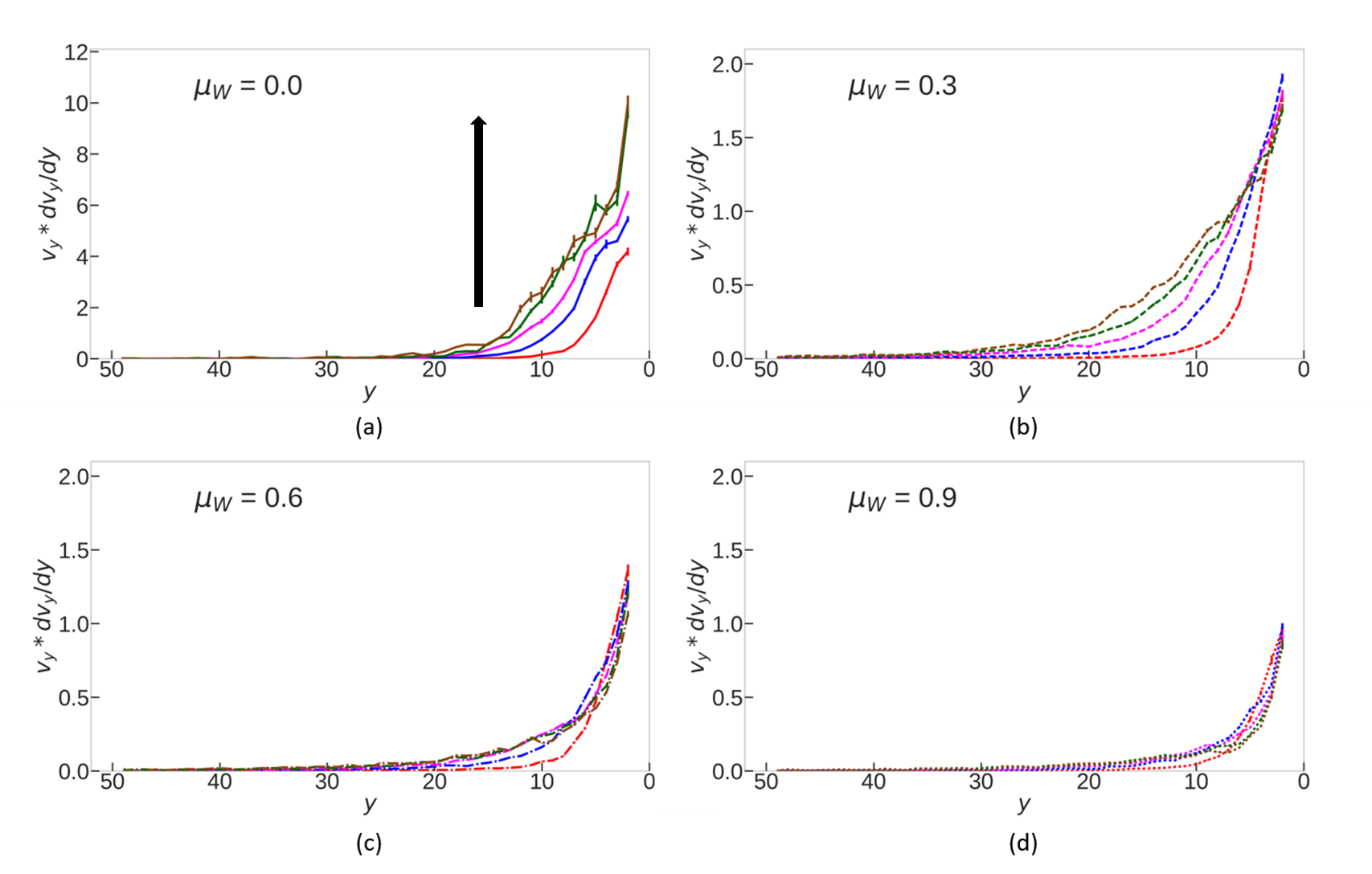}
	\setcounter{figure}{11}
	\caption{This figure shows the acceleration of the particles along the central line of the hopper for the different wall $\mu$ and varying opening sizes. The arrow indicates the direction of increasing opening size, and is representative for all wall $\mu$. The accelerations are found by binning the central vertical line of the hopper with $1 \times 1$ boxes, in which we average $v_y \frac{d v_y}{dy}$ over time and initializations.}
	\label{Accelerations}
\end{figure*}

\newpage

\end{document}